\documentclass[conference]{IEEEtran}
\usepackage{graphicx}
\usepackage{latexsym}
\usepackage{amsmath}
\usepackage{url}
\begin{document}
\title{How to Make Chord Correct}
\author{\IEEEauthorblockN{Pamela Zave}
\IEEEauthorblockA{AT\&T Laboratories---Research\\
Bedminster, New Jersey 07921, USA\\
Email: pamela@research.att.com}}
\maketitle

\begin{abstract}
The Chord distributed hash table (DHT) is well-known and frequently used to
implement peer-to-peer systems.
Chord peers find other peers, and access their data,
through a ring-shaped pointer
structure in a large identifier space.
Despite claims of proven correctness, i.e., eventual reachability,
previous work
has shown that the Chord ring-maintenance protocol is not correct
under its original operating assumptions.
It has not, however, discovered whether Chord could be made correct
with reasonable operating assumptions.
The contribution of this paper is to provide the first specification
of correct operations and initialization
for Chord, an inductive invariant that is necessary and
sufficient to support a proof of correctness, and the proof itself.
Most of the proof is carried out by automated analysis of an Alloy model.
The inductive invariant reflects the fact that a Chord network must
have a minimum ring size (the minimum being the length of successor lists
plus one) to be correct.
The invariant relies on an assumption
that there is a stable base, of the minimum size,
of permanent ring members.
Because a stable base has only a few members and a Chord network can
have millions, we learn that the obstacles to provable correctness
are anomalies in small networks, and that a stable base need not be
maintained once a Chord network grows large.
\end{abstract}

\section{Introduction}

Peer-to-peer systems are distributed systems featuring decentralized
control, self-organization of similar nodes, 
and scalability.
A distributed hash table (DHT) is a peer-to-peer system that implements
a persistent key-value store.
It can be used for shared file storage, group directories, and many
other purposes.

The distributed hash table Chord 
was first presented in a 2001 
SIGCOMM paper \cite{chord-sigcomm}.
This paper was the fourth-most-cited paper in computer science for
several years (according to Citeseer), and won
the 2011 SIGCOMM Test-of-Time Award.

The nodes of a Chord network have identifiers in an {\it m}-bit
identifier space, and
reach each other through pointers in this identifier space.
Because the pointer structure is based on adjacency in the identifier
space, and $2^{m} - 1$ is adjacent to 0, the structure of a
Chord network is a ring.

The ring structure is disrupted when nodes join, leave, or fail.
The original Chord papers \cite{chord-sigcomm,chord-ton}
specify a ring-maintenance protocol whose 
minimum correctness property is eventual
reachability:  given ample time and no further disruptions, 
the ring-maintenance protocol can repair all disruptions in the ring
structure.
If the protocol is not correct in this sense, then some nodes of a
Chord network will become permanently unreachable from other nodes.

The introductions of the original Chord papers
say, ``Three features that
distinguish Chord from many other peer-to-peer lookup protocols
are its simplicity, provable correctness, and provable performance.''
An accompanying PODC paper \cite{chord-podc}
lists invariants of the ring-maintenance protocol.

The claims of simplicity and performance are certainly true.
The Chord algorithms are far simpler and more completely specified
than those of other DHTs, such as Pastry \cite{pastry},
Tapestry \cite{tapestry}, CAN \cite{CAN}, and Kademlia \cite{kademlia}.
There is no attempt to specify synchronization or timing constraints
on distributed nodes.
There are no atomic operations involving multiple nodes.

The ease of implementing Chord is probably the reason for its
popularity as a component of peer-to-peer systems.
Its fundamental simplicity is probably the reason for its
popularity as a
basis for building DHTs with stronger guarantees and additional
capabilities, such as
protection against malicious peers
\cite{awerbuch-robust,chord-byz,sechord},
key consistency and data consistency \cite{scatter},
range queries \cite{rangequeries},
and atomic access to replicated data \cite{atomicchord,etna}.

Unfortunately, the claim of correctness is not true.
The original specification with its original operating
assumptions does not have eventual reachability,
and {\it not one} of the seven properties claimed to be invariants
in \cite{chord-podc} is actually an invariant
\cite{chord-ccr}.
This was revealed by modeling the protocol in the Alloy language
and checking its properties with the Alloy Analyzer \cite{alloy-book},
an exercise that illustrates rather clearly the importance of formal
modeling of protocols.

The principal contribution of this paper
is to provide the first specification of a
version of Chord that is correct under reasonable operating
assumptions.
It corrects all the flaws that were revealed in \cite{chord-ccr},
as well as some new ones.
In addition, the paper provides a concise, necessary, and sufficient
inductive invariant.
It also provides the proof of correctness.

It has been said, of the flaws in original Chord, that they are
either obvious and fixed by all implementers,
or extremely unlikely to cause trouble during Chord execution.
Taking this comment into account, 
the results in this paper are significant in the
following ways:

(1)
Many people implement Chord, or
use Chord as a component of their distributed systems.
At least some of them do not discover the flaws in original Chord 
{\it e.g.,} \cite{overlog}.
Implementers should have a correct version of Chord to use, and they
should not have to discover it for themselves.
They should also know the invariant for Chord, as dynamic checking
of the invariant is a
design principle for enhancing DHT security \cite{sitmorris}.

(2) There is no way to know that the scenarios of subtle bugs 
are truly improbable, for all implementations.
To estimate the probabilities, it would be necessary to make
a number of assumptions about implementation-specific attributes
such as timing.

(3)
Many people build on Chord, and reason about Chord behavior,
for the purposes of their research.
This reasoning should have a sound foundation. 
For example,
the performance analysis in \cite{chord-churn} makes incorrect
assumptions about Chord behavior \cite{chord-ccr}.
The research on augmenting and strengthening Chord, as referenced above,
relies on informal descriptions of Chord and informal reasoning about
its behavior.
As automated proof checking increasingly 
becomes the norm in distributed systems,
attempts to prove properties of systems based on original Chord
will fail or yield unsound results.
Most automated reasoning is absolute rather than probabilistic,
so even improbable bugs would make it unsound.

(4)
As will be explained in Section~\ref{sec:additional}, efforts to find the
best version of Chord and the best invariant for a proof have led to
interesting insights into how Chord works.
People who build on Chord should be aware of these properties so as to
preserve them and to benefit from them.
In one example given in Section~\ref{sec:stable}, the proof shows that
Chord can be implemented more efficiently than was originally believed.
Some principles may be applicable to all systems that use ring-shaped
pointer structures in large identifier spaces 
({\it e.g.,} \cite{awerbuch-hyperring,CAN}).

The paper begins with an overview of Chord using the revised, correct
ring-maintenance operations
(Section~\ref{sec:overview}), and a new specification of these
operations (Section~\ref{sec:spec}).
Although the specification
is pseudocode for immediate accessibility, it is a paraphrase
of a formal specification in Alloy.
The complete Alloy model, including specification, invariant, and 
all steps of the proof,
can be found at
\url{http://www2.research.att.com/~pamela/chord.html}.
In addition, Section~\ref{sec:diff} provides a brief summary of 
differences between the original and correct Chord operations,
and why the differences matter.

Correct operations are necessary but not sufficient.
It is also necessary to have an inductive invariant
to use in constructing a proof, and to
initialize a network in a state that satisfies the invariant.
Original Chord is initialized with a network of one node,
which is
not correct, and
Section~\ref{sec:additional} shows why.
Chord must be initialized 
with a ring containing a minimum of $r + 1$ nodes, where
$r$ is the length of each node's list of successors.

In fact, to be proven correct, a Chord network must maintain a
``stable base'' of $r + 1$ nodes that
remain members of the network throughout its lifetime.
Section~\ref{sec:additional} shows that
a stable base enforces a concise invariant that implies
other necessary structural properties.
The section also explains that, while a stable base is necessary for
provable correctness, the anomalies it is preventing can only occur
in small rings.
Thus, when the ring is large, a stable base need not be maintained.

The proof in
Section~\ref{sec:proof}
has both manual and automated parts.
The automated parts establish the invariant and guarantee that, if
the state of the network is non-ideal, some repair
operation is enabled that
will change the network state.
The manual part defines a measure, which is a non-negative integer,
of the error in a non-ideal network.
It also shows that every state change due to a repair operation reduces
the error.
Together these parts show that if all enabled operations
occur eventually, then repair operations will eventually reduce
the error to zero, at which time the network state will be ideal.

\begin{figure}
\centering
\includegraphics[scale=0.80]{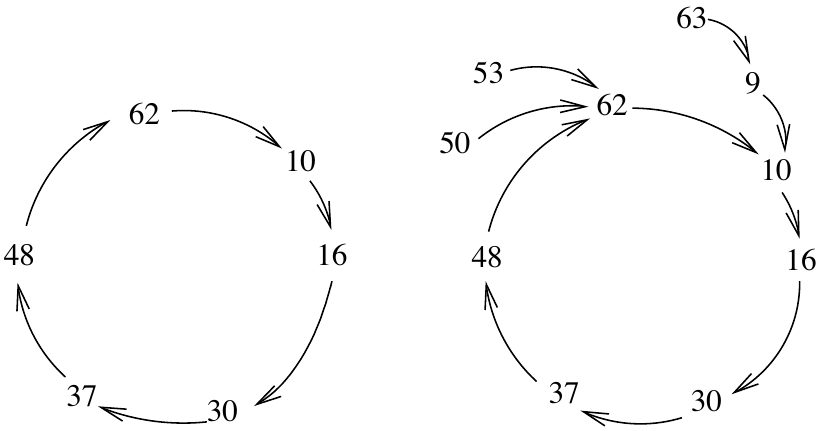}
\caption{Ideal (left) and valid (right) networks.
Members are represented by their identifiers.
Solid arrows are successor pointers.}
\label{fig:valid}
\end{figure}

The conclusion
(Section~\ref{sec:proof}) includes recommendations for implementers
and future work.

Together Sections~\ref{sec:diff} and
\ref{sec:additional} present most of the problems with original
Chord reported
in \cite{chord-ccr} (as well as previously unreported ones).
The problems are not presented first because they make more sense when
explained along with their underlying nature and how to remove them.

Although other researchers have found problems with Chord implementations
\cite{chord-nontrans,mace,crystalball}, they have not discovered any
problems with the specification of Chord.
Other work on verifiable ring maintenance operations \cite{ringtop}
uses multi-node atomic operations, which are avoided by Chord.

\section{Overview of correct Chord}
\label{sec:overview}

Every member of a Chord network has an identifier (assumed unique) that
is an {\it m}-bit hash of its IP address.
Every member has a {\it successor list} of pointers to other members.
The first element of this list is the {\it successor}, and is
always shown as a solid
arrow in the figures.
Figure~\ref{fig:valid} shows two Chord networks with {\it m} = 6,
one in the ideal state of a ring ordered by identifiers,
and the other
in the valid state of an ordered ring with appendages.
In the networks of Figure~\ref{fig:valid}, key-value pairs with keys
from 31 through 37 are stored in member 37.
While running the ring-maintenance protocol, a member also acquires and
updates a {\it predecessor} pointer, which is always shown as a dotted
arrow in the figures.

\begin{figure*}
\centering
\includegraphics[scale=0.80]{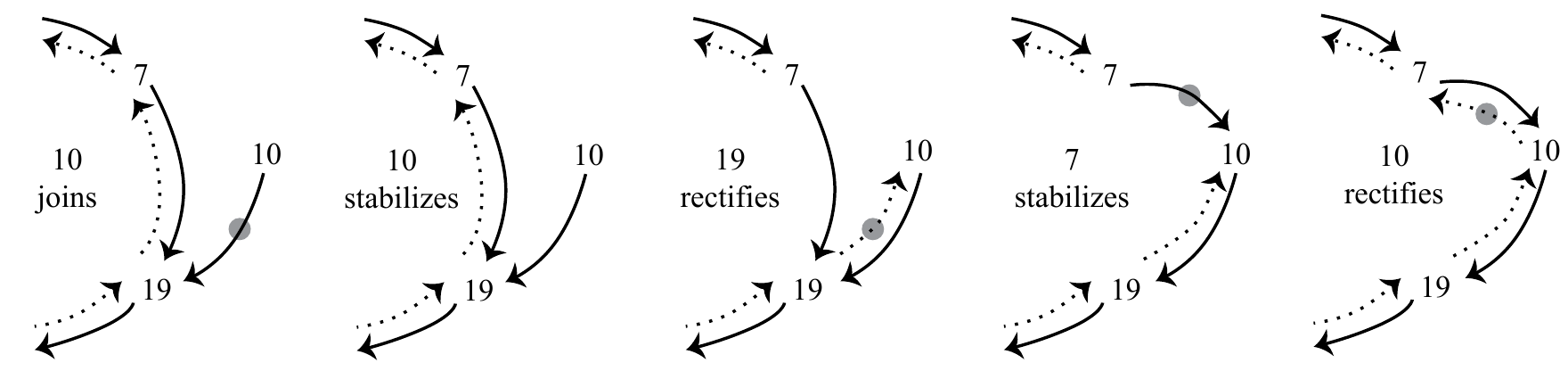}
\caption{A new node becomes part of the ring.
A gray circle marks the pointer updated by an operation, if any.
Dotted arrows are predecessors.}
\label{fig:join}
\end{figure*}

The ring-maintenance protocol is specified in terms of three
operations, each of
which changes the state of at most one member.
In executing an operation, the member queries another member or
sequence of members, then
updates its own pointers if necessary.
The specification of Chord assumes that inter-node communication is
bidirectional and
reliable, so we are not concerned with Chord behavior when inter-node
communication fails.

A node becomes a member in a {\it join} operation.
A member node is also referred to as {\it live}.
When a member joins, it contacts an existing member and gets its own
current successor from that member.
(It also contacts the current successor to get a full successor list.)
The first stage of Figure~\ref{fig:join} shows successor and predecessor
pointers in a section of a network where 10 has just joined.

When a member {\it stabilizes}, it learns its successor's predecessor.
It adopts the predecessor
as its new successor, provided that the predecessor
is closer in identifier order
than its current successor.
Because a member must query its successor to stabilize, this is also
an opportunity for it to update its successor list with information
from the successor.
Members schedule their own stabilize operations, which should
be periodic.

Between the first and second stages
of Figure~\ref{fig:join}, 10 stabilizes.
Because its successor's predecessor is 7, which is not a better successor
for 10 than its current 19, this operation does not change the successor
of 10.

After stabilizing (regardless of the result),
a node notifies its successor of its identity.
This causes the notified member to execute a {\it rectify} operation.
The rectifying member checks whether its current predecessor is still a
member, and then adopts the notifying member as its new predecessor
if the notifying member
is closer in identifier order than its current predecessor
(or if it has no live predecessor).
In the third stage of Figure~\ref{fig:join},
10 has notified 19, and 19 has adopted 10 as its new predecessor.

In the fourth stage of Figure~\ref{fig:join}, 7 stabilizes, which causes
it to adopt 10 as its new successor.
In the last stage 7 notifies and 10 rectifies, 
so the predecessor of 10 becomes 7.
Now the new member 10 is completely incorporated into the ring, and all
the pointers shown are correct.

One operating
assumption of the protocol is that a member in good standing always
responds to queries in a timely fashion.
A node ceases to become a member in a {\it fail} event, which can
represent failure of the machine, or the node's
silently leaving the network.
A member that has failed is also referred to as {\it dead}.
Another operating assumption is that, after a member fails, it no longer
responds to queries from other members.
With these assumptions, members can detect the failure of other members
perfectly by noticing whether they respond to a query before a timeout
occurs.
A third assumption about failure behavior is that
successor lists are long enough, and failures are infrequent enough,
to ensure that a
member is never left with no live successor in its list.

Failures can produce gaps in the ring, which are repaired during
stabilization.
As a member attempts to query its successor for stabilization, 
it may find that its
successor is dead.
In this case it attempts to query the next member in its successor
list and make this its new successor, continuing through the list
until it finds a
live successor.

As in the original Chord papers \cite{chord-sigcomm,chord-ton},
we wish to define a correctness property of eventual reachability:
given ample time and no further disruptions, 
the ring-maintenance protocol can repair disruptions so that
every member of a Chord network is reachable from every other member.
Note that a network with appendages (nodes 50, 53, 63, 9 on the right
side of Figure~\ref{fig:valid}) cannot have full reachability,
because an appendage cannot be reached by a member that is not an
appendage.

A network is {\it ideal} when each pointer is globally correct.
For example, on the right of Figure~\ref{fig:valid}, the globally correct
successor of 48 is 50 because it is the nearest member in identifier
order.
Because the ring-maintenance protocol is supposed to repair all
imperfections, and because it is given ample time to do all
the repairs, the correctness criterion can be strengthened
slightly, to:
{\it In any execution state, if there are no
subsequent join or fail events, then eventually the network will become
ideal and remain ideal.}

Defining a member's {\it best successor} as its first successor pointing
to a live node (member), a {\it ring member} is a member that can reach
itself by following the chain of best successors.
An {\it appendage member} is a member that is not a ring member.
Of the seven invariants presented in
\cite{chord-podc} (and all violated by original Chord),
the following four are necessary for correctness.
\begin{itemize}
\item
There must be a ring, which means that there must be a non-empty set
of ring members ({\it AtLeastOneRing}).
\item
There must be no more than one ring, which means that from each ring
member, every other ring member is reachable by following the chain
of best successors ({\it AtMostOneRing}).
\item
On the unique ring, the nodes must be in identifier 
order ({\it OrderedRing}).
\item
From each appendage member, the ring must be reachable by following
the chain of
best successors ({\it ConnectedAppendages}).
\end{itemize}
If any of these rules is violated,
there is a disruption in the structure that
the ring-maintenance protocol cannot repair, and some members
will be permanently unreachable from some other members.
It follows that any inductive invariant for Chord must include these
as conjuncts.

The Chord papers define the lookup protocol, which is not discussed here.
They also define the maintenance and use of finger tables, which 
improve lookup speed
by providing pointers that cross the ring like
chords of a circle.
Because finger tables are an optimization and 
they are built from successors and predecessors, correctness
does not depend on them.

\section{Specification of ring-maintenance operations}
\label{sec:spec}

This section contains pseudocode, derived from the Alloy model,
for the join, stabilize, and rectify operations.

There is a type 
\small
{\tt Identifier}
\normalsize
which is a string of {\it m} bits.
Implicitly, whenever a member transmits the identifier of a member, it
also transmits its IP address so that the recipient can reach the 
identified member.
The pair is self-authenticating, as the identifier must be the hash of the
IP address according to a chosen function.

The Boolean function 
\small
{\tt between}
\normalsize
is used to check the order of
identifiers.
Because identifier order wraps around at zero,
it is meaningless to compare two identifiers---each precedes
and succeeds the other.
This is why 
\small
{\tt between}
\normalsize
has three arguments:
\small
\begin{verbatim}
Boolean function between (n1,n2,n3: Identifier) 
{  if (n1 < n3) return ( n1 < n2 && n2 < n3 )
   else         return ( n1 < n2 || n2 < n3 )     
}
\end{verbatim}
\normalsize
It is important to note that, for all distinct 
\small
{\tt x}
\normalsize
and 
\small
{\tt y}, {\tt between(x,y,x)}
\normalsize
is always true, 
and 
\small
{\tt between(x,x,y)}
\normalsize
and
\small
{\tt between(y,x,x)}
\normalsize
are always false.

The function
\small
\begin{verbatim}
Identifier function lookupSucc 
   (joining: Identifier) { }
\end{verbatim}
\normalsize
takes the identifier of a joining node, and uses the lookup
protocol to return the identifier of
its proper successor in the ring.
In other words, for two members 
\small
{\tt n}
\normalsize
and
\small
{\tt lookupSucc(joining)}
\normalsize
that are adjacent in the ring,
\small
{\tt between(n,joining,lookupSucc(joining)).}
\normalsize

Each node has the following variables:
\small
\begin{verbatim}
myIdent: Identifier;
known: Identifier;
pred: Identifier U Null;
succList: list Identifier;      // length is r
\end{verbatim}
\normalsize
where
\small
{\tt myIdent}
\normalsize
is the hash of its IP address,
\small
{\tt known}
\normalsize
is a member of the Chord network known to the node
when it joins, and
\small
{\tt pred}
\normalsize
is the node's predecessor.
For convenience in the pseudocode, we allow the type
\small
{\tt Identifier}
\normalsize
to include the constant
\small
{\tt Null},
\normalsize
meaning that there is no predecessor.
\small
{\tt succList}
\normalsize
is its entire successor list; the head of this list
is its {\it first successor} or simply its {\it successor}.
The parameter {\it r} is the fixed length of all successor lists.

To join, a node executes the following pseudocode.
\small
\begin{verbatim}
// Join operation

newSucc: Identifier;

query known for lookupSucc(myIdent);
if (query returns before timeout) {
   newSucc = lookupSucc(myIdent);
   query newSucc for newSucc.succList;
   if (query returns before timeout) {
      succList =
         append(newSucc,
                butLast(newSucc.succList));
      pred = Null;
   }
   else retry Join later;
}
else retry Join later;
\end{verbatim}
\normalsize
First, the node asks the known node to look up the node's identifier and
get its proper successor, storing the value in
\small
{\tt newSucc}.
\normalsize
The node then queries
\small
{\tt newSucc}
\normalsize
for its successor list.
Finally the node constructs its own successor list by
concatenating
\small
{\tt newSucc}
\normalsize
and
\small
{\tt newSucc}'s
\normalsize
successor list, with the
last element of the list trimmed off to produce a
result of length {\it r}.
If either of the queries fail the node has no choice but to retry
again later.

To stabilize, a node executes the following pseudocode.
\small
\begin{verbatim}
// Stabilize operation

newSucc: Identifier;

while (succList is not empty) {
   query head(succList) for 
      head(succList).pred and
      head(succList).succList;
   if (query returns before timeout) {
      newSucc = head(succList).pred;
      succList =
         append(
            head(succList),
            butLast(head(succList).succList)
         );
      if 
      (between(myIdent,newSucc,head(succList))
      {  query newSucc for newSucc.succList;
         if (query returns before timeout) 
            succList =
               append(
                  newSucc,
                  butLast(newSucc.succList)
               );
      };
      notify head(succList) of myIdent;
      break;
   }
   else succList = tail(succList);
};
\end{verbatim}
\normalsize
In the outer loop of this code, the
node queries its successor for its successor's predecessor and
successor list.
If this query times out, then the node's successor is presumed dead.
The node promotes its second successor to first and tries again.
Once it has contacted a live successor, it executes inner code ending
in a break out of the loop.
The loop is guaranteed to terminate before
\small
{\tt succList}
\normalsize
is empty,
based on the assumption that
successor lists are long enough so that each list contains at least one
live node.

Once it has contacted a live successor, the node first updates
its successor list with its successor's list.
It then checks to see if the new pointer it has learned, its successor's
predecessor, is an improved successor.
If so, and if
\small
{\tt newSucc}
\normalsize
is live, it adopts
\small
{\tt newSucc}
\normalsize
as its new successor.
Thus the stabilize operation requires one or two queries for each
traversal of the outer loop.
Whether or not there is a live improved successor, the node notifies
its successor of its own identity.

A node rectifies when it is notified, thereafter executing the following
pseudocode:
\small
\begin{verbatim}
// Rectify operation

newPred: Identifier;

receive notification of newPred;
if (pred = Null) pred = newPred;
else {
   query pred to see if live;
   if (query returns before timeout) {
      if (between(pred,newPred,myIdent))
         pred = newPred;
   }
   else pred = newPred;
};
\end{verbatim}
\normalsize

When a node fails or leaves,
it ceases to stabilize, notify, or respond to queries
from other nodes.
When a node rejoins, it re-initializes its Chord variables.

\section{Differences between the versions}
\label{sec:diff}

The {\it join, stabilize,} and {\it notified}
operations of the original protocol are defined as pseudocode in
\cite{chord-sigcomm} and \cite{chord-ton}.
These papers do not provide details about failure recovery.
The only published paper with pseudocode for failure recovery is
\cite{chord-podc}, where failure recovery is
performed by the
{\it reconcile, update,} and {\it flush} operations.
The following table shows how events of the two versions correspond.
Although {\it rectify} in the new version is similar to
{\it notified} in the old version, it seems more consistent to use
an active verb form for its name.

\begin{center}
\footnotesize
\begin{tabular}{|c|c|}
\hline
{\bf old} & {\bf new} \\ \hline
join +	& join \\
reconcile &  \\ \hline
stabilize + & stabilize \\
reconcile + & \\
update & \\ \hline
notified + & rectify \\
flush & \\ \hline
\end{tabular}
\normalsize
\end{center}

In both old and new versions of Chord, members schedule their own
maintenance operations except for {\it notified} and
{\it rectify}, which occur when a member is notified by its 
predecessor.
Although the operations
are loosely expected to be periodic, scheduling is not formally
constrained.
As can be seen from the table, 
multiple smaller operations from the old version
are assembled into larger new operations.
This ensures that the successor lists of members are always fully
populated with $r$ entries,
rather than having missing entries to be filled in by later
operations.
An incompletely populated successor list might lose (to failure)
its last live
successor.
If the successor list belongs to an appendage member, this would mean
that the appendage can no longer reach the ring, which is
a violation of {\it ConnectedAppendages} \cite{chord-ccr}.

Another systematic change from the old version to the new is that,
before incorporating a pointer to a node into its state, a
member checks that it is live.
This prevents cases where a member replaces a pointer to a live node with
a pointer to a dead one.
A bad replacement can also cause a successor list to have no live
successor.
If the successor list belongs to a ring member, this will cause a
break in the ring, and a 
violation of {\it AtLeastOneRing}.
Together these two systematic changes also prevent
scenarios in which the ring becomes disordered or breaks into
two rings of equal size
(violating {\it OrderedRing} or {\it AtMostOneRing}, respectively
\cite{chord-ccr}).

A third systematic change is that the new code
is much more complete and explicit than the original
pseudocode, particularly with respect to communication
between nodes.
This is important because a Chord operation at a single node
can entail multiple
queries to other nodes.
Thus the operation has multiple phases that can be interleaved with
operations at other nodes, and the proof of correctness must consider
these interleavings.

In addition to these systematic changes, a few other small problems
were detected by Alloy modeling and analysis, and fixed.

\begin{figure*}
\centering
\includegraphics[scale=0.80]{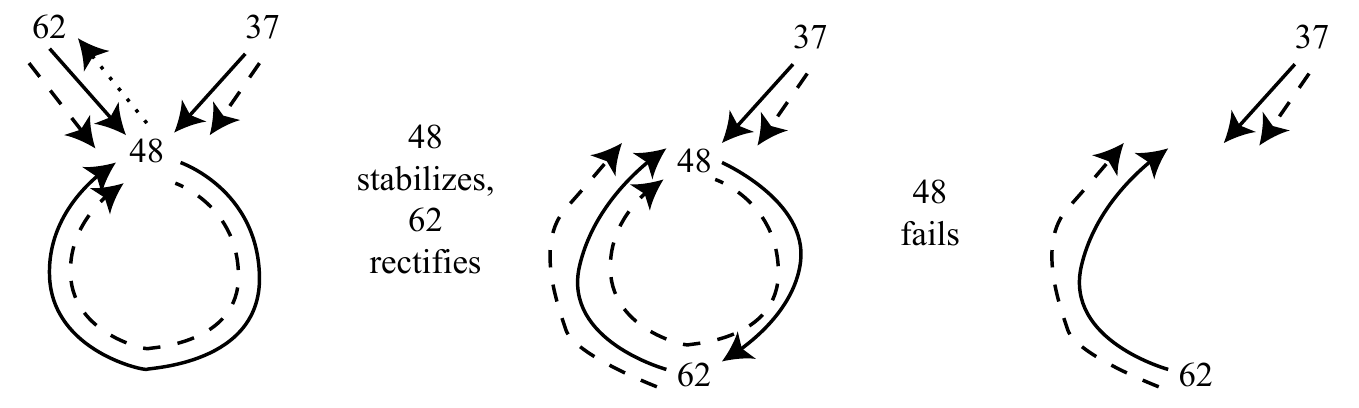}
\caption{Why the ring cannot be initialized at size 1.
Dashed arrows are second-successor pointers.
Predecessor pointers are not shown in the last two stages, as they are
irrelevant.
This problem was not reported in \cite{chord-ccr}.}
\label{fig:init}
\end{figure*}

\begin{figure*}
\centering
\includegraphics[scale=0.80]{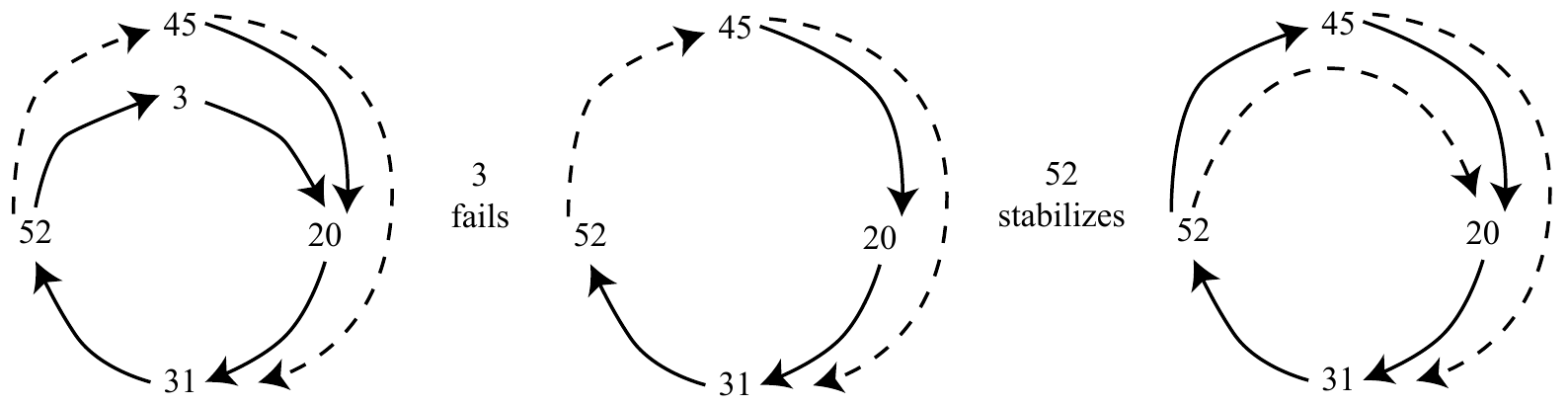}
\caption{A counterexample to a trial invariant.
Only the relevant pointers are drawn.}
\label{fig:wrap}
\end{figure*}

\section{The inductive invariant}
\label{sec:additional}

An {\it inductive invariant} is an invariant with the property that
if the system satisfies the invariant before any action or event,
then the system can be proved to satisfy the invariant after the 
action or event.
By induction, if the system's initial state satisfies the invariant,
then all system states satisfy the invariant.
Typically an inductive invariant is a conjunction of non-inductive
invariants, each of which is not strong enough by itself to be inductive.

Correct operations for Chord are necessary but not sufficient.
We also need an inductive invariant to use in
constructing a proof, and the network must be initialized to a state
that satisfies the invariant.

This section will describe concepts in terms of a node's
{\it extended successor list,}
which is simply its successor list with the node identifier in 
front.
So node
\small
{\tt n}'s
\normalsize
extended successor list, of size $r + 1$, is
\small
{\tt append(n, n.succList).}
\normalsize
Conjuncts of the inductive invariant
are all defined formally in Alloy, and will be
presented as informal paraphrases here.

\subsection{Minimum size}

Of the four conjuncts defined in Section~\ref{sec:overview},
three of them constrain the network to have a single ordered ring
({\it AtLeastOneRing, AtMostOneRing,} and {\it OrderedRing}), while
{\it ConnectedAppendages} constrains the appendage members to be able to
reach the ring.
All are necessary in the invariant,
but even together are not sufficient.

As stated previously, a node's 
successor list must have $r$ entries because this is necessary
to guarantee, under the protocol's operating assumptions, that the node
will always have a live successor.
For the same reason, each extended successor list must have 
$r + 1$ {\it distinct} entries.

Original Chord initializes a network with a single member that is its
own successor, {\it i.e.,} the initial network is a ring of size 1.
This is not correct, as shown in
Figure~\ref{fig:init} with $r = 2$.
Appendage nodes 62 and 37 start with both list entries equal to 48.
Then 48 fails, leaving members 62 and 37
with insufficient information to find each other.
For members to be able to
have $r + 1$ distinct entries in ideal extended successor lists,
a Chord network must be initialized and maintained with a
minimum ring size of $r + 1$.

It seems that we need a conjunct {\it NoDuplicates} stating that a
node's extended successor list has no duplicated entries.
This implies a minimum ring size, but it is impossible to enforce
with normal Chord operations.
Chord operations are local, and a member does not know how many other
members or ring members there are.
We will return to this issue in the next section.

\subsection{Preventing disorder}
\label{sec:disorder}

Because a node's successor list is ideally intended to replicate
and/or become the ring structure, it seems wise to have
a conjunct
{\it OrderedSuccessorLists} saying that
for all contiguous sublists
\small
{\tt (x, y, z)}
\normalsize
of
a node's extended successor list,
\small
{\tt between(x, y, z)}
\normalsize
holds.

Unfortunately the four original conjuncts and the two new conjuncts
{\it NoDuplicates} and {\it OrderedSuccessorLists}
are still not
sufficient to provide an inductive invariant.
To give one of a multitude of counterexamples, consider
Figure~\ref{fig:wrap}, which is another example with $r = 2$.
The first stage satisfies the trial invariant, having duplicate-free
and ordered extended
successor lists such as (52, 3, 45) and (45, 20, 31).
The appendage node 45 does not merge into the ring at the correct
place, but that is part of normal Chord operation (see \cite{chord-ccr}).
The second successor of ring node 52 points outside the ring, but
that is also part of Chord operation (see Appendix A).
Once 3 fails and 52 stabilizes, however, 
the ring becomes disordered.

There is a stronger invariant that allows Chord to be proved correct.
It relies on an operating assumption that a Chord network is initialized
with a set of members containing a {\it stable base} of at least $r + 1$
members.
The typical range for {\it r} is 3-5, so the typical stable base would
require 4 to 6 members.
These members are ``stable'' in the sense that they continue to be
members throughout the life of the network, without ever leaving or
failing and rejoining.
Because a member's identifier is derived from its IP address, this
means that there is always a live IP host at that address, with
a copy of the member state for that identifier.

The remainder of this section will explain the invariant supported
by the assumption of a stable base, and how it provides the structure
needed to prove that Chord is correct.
Section~\ref{sec:stable} discusses the stable base further, answering
the two key questions of what it means for implementers and why it
is a necessary assumption to prove correctness.

The final inductive invariant is the conjunction of 
{\it AtLeastOneRing, AtMostOneRing, OrderedRing, ConnectedAppendages,}
and {\it BaseNotSkipped}.
To explain {\it BaseNotSkipped},
we say that a member
\small
{\tt n}
\normalsize
{\it skips} a member
\small
{\tt n2}
\normalsize
if there is an adjacent pair 
\small
{\tt (n1, n3)}
\normalsize
in the extended successor list of
\small
{\tt n},
\normalsize
and
\small
{\tt between(n1, n2, n3)}
\normalsize
(which implies that neither
\small
{\tt n1}
\normalsize
nor
\small
{\tt n3}
\normalsize
is
\small
{\tt n2}).
\normalsize
A member
\small
{\tt n}
\normalsize
typically skips
\small
{\tt n2}
\normalsize
if
\small
{\tt n2}
\normalsize
became a
member recently, so that knowledge of it has not yet reached 
\small
{\tt n}.
\normalsize
{\it BaseNotSkipped} says that no member
of a Chord network skips a member of the stable base.
{\it BaseNotSkipped} excludes the first stage of Figure~\ref{fig:wrap},
because 52 skips 20 and 31.
Of the four ring members 3, 20, 31, and 52, at least three must be in
the stable base, so 52 cannot skip two of them and still satisfy
the invariant.

We can reason directly about how {\it BaseNotSkipped} prevents
counterexamples such as the one in Figure~\ref{fig:wrap}.
A counterexample network has two extended successor lists
\small
{\tt (x,} 
\normalsize
{\it failing\_nodes,}
\small
{\tt y, . . . )}
\normalsize
and 
\small
{\tt (y, . . . z,}
\normalsize
{\it at\_least\_one\_node})
where 
\small
{\tt between(x, z, y)}.
\normalsize
When the failing nodes fail and 
\small
{\tt x}
\normalsize
stabilizes, the extended successor list of 
\small
{\tt x}
\normalsize
becomes 
\small
{\tt (x, y, . . . z, . . .)}
\normalsize
which ``wraps around'' the ring if it is interpreted as a clockwise
path---once the path has reached 
\small
{\tt z},
\normalsize
it has passed its origin 
\small
{\tt x}.
\normalsize

Can this counterexample be constructed and still satisfy 
{\it BaseNotSkipped}?
The trick is to fit the $r + 1$ base nodes into the extended successor
lists.
\small
{\tt x}
\normalsize
and
\small
{\tt y}
\normalsize
can be base nodes but
\small
{\tt z}
\normalsize
cannot, because it would be skipped by the extended successor list of
\small
{\tt x}.
\normalsize
To satisfy {\it BaseNotSkipped}, the remaining $r - 1$ base nodes
must fit into the ellipsis between 
\small
{\tt y} 
\normalsize
and
\small
{\tt z}
\normalsize
in the extended successor list of
\small
{\tt y}.
\normalsize
This is not possible, however, because the length of the extended
successor list is $r + 1$, so the maximum length of the ellipsis is
$r - 2$.

This argument demonstrates how {\it BaseNotSkipped} serves to fill out
successor lists so that they do not span too big an arc of the ring.
It is easy to see that {\it BaseNotSkipped} implies
{\it NoDuplicates}.
If an extended successor list mentions node 
\small
{\tt n}
\normalsize
twice, then even if
\small
{\tt n}
\normalsize
is a base node, the other $r$ base nodes must fit in the space between
the two mentions, or otherwise they would be skipped.
Yet the maximum size of the space is $r - 1$.

Although it is a little harder to see, {\it BaseNotSkipped}
also implies {\it OrderedSuccessorLists}.
Here is an informal
proof by contradiction:

Contrary to the hypothesis, assume there is an extended successor
list of the form
\small
{\tt ( . . . x, z, y, . . . )}
\normalsize
where 
\small
{\tt between(x, z, y)}
\normalsize
is false.
This means that a clockwise path around the ring from
\small
{\tt x}
\normalsize
would go through 
\small
{\tt y},
\normalsize
then
\small
{\tt z},
\normalsize
then come to
\small
{\tt x}
\normalsize
again.

Since the extended successor list must satisfy {\it BaseNotSkipped},
we can ask where the base nodes are in the ring.
\begin{itemize}
\item
There cannot be base nodes in the arc of the ring between
\small
{\tt x}
\normalsize
and 
\small
{\tt z},
\normalsize
which includes
\small
{\tt y},
\normalsize
because the pair
\small
{\tt (x, z)}
\normalsize
would skip them.
\item
There cannot be base nodes in the arc of the ring between
\small
{\tt z}
\normalsize
and
\small
{\tt y},
\normalsize
which includes
\small
{\tt x},
\normalsize
because the pair
\small
{\tt (z, y)}
\normalsize
would skip them.
\end{itemize}
Thus the arcs where base nodes are prohibited cover the entire ring
except
\small
{\tt z}.
\normalsize
A stable base always has more than one member, so there is a contradiction.

It has now been shown that
adding {\it BaseNotSkipped}
to the four original conjuncts
prevents duplicates in successor lists and guarantees ordered
successor lists.
Thus {\it BaseNotSkipped} is a powerful and surprisingly
compact representation
of the structure of a correct Chord network.

\section{Discussion of the stable base}
\label{sec:stable}

\subsection{What does a stable base mean for implementers?}

There is something rather odd about the assumption of a stable base:
a stable base has few nodes and a Chord network can have millions.
Furthermore, there is no requirement on how members of the
stable base are distributed around the ring.
This means that there are arbitrarily large sections of the ring that
are not close to any member of the stable base, and whose operations
have nothing to do with the stable base.
So how is the stable base maintaining their correctness?

The solution to this puzzle is obvious in retrospect.
Once the operations are made correct
as described in Sections~\ref{sec:spec}
and \ref{sec:diff}, 
all the remaining correctness problems arise from anomalies in
small rings:
they fall below minimum size, their successor lists ``wrap around''
the ring, {\it etc.}
Once the ring has grown large, it is not in danger of any such
anomalies, and the stable base is not needed.

This is good news for implementers.
For a correct implementation of Chord, it is necessary to initialize
it with $r + 1$ members, and to preserve a stable base until the
network grows to a safe size (perhaps three times the size of the
stable base).
After that, provided that the network does not shrink drastically,
the stable-base assumption can be ignored.

There is an additional bonus for implementers.
Consider what happens when a
member node fails, recovers, and
wishes to rejoin, all of which could occur within a short period of
time.
It was previously thought necessary
for the node to wait until all
previous references to its identifier had probably been 
cleared away, because obsolete pointers could be incorrect in the
current state. 
This wait was included in the first Chord
implementation \cite{excuses}.
Yet the stable base makes the wait unnecessary, as Chord is provably
correct even with obsolete pointers.

In the spirit of \cite{sitmorris}, it is a good security practice
to monitor that invariants are satisfied.
All of the conjuncts of the inductive invariant are global, and thus
unsuitable for local monitoring.
The right properties to monitor are {\it NoDuplicates} and
{\it OrderedSuccessorLists}, which
can be checked on individual successor lists.
These are the properties that must be invariant in Chord networks of
any size.

\subsection{Is a stable base necessary to prove correctness?}

There is strong evidence that {\it BaseNotSkipped} is a necessary
conjunct in the inductive invariant of Chord.
We have seen that it implies other necessary properties.
Most importantly,
it guarantees that the network maintains its minimum size, which
would not otherwise be possible without additional coordination
mechanisms.

The final argument is that there was an enormously long and
ultimately fruitless search for another invariant.
An inductive invariant is a concise characterization of all the states
that a system can reach during operation, even if it is operating
for a long time.
To prove that it is an inductive invariant, we must show that every
system operation preserves it.
Consequently, there are two requirements for an inductive invariant: 
(1) it must be strong enough so that every operation has a well-structured
state to work on;
(2) it must be weak enough so that the result of every operation still
satisfies it.
Unfortunately these two requirements conflict, making inductive
invariants notoriously difficult to find.

For those who would like a taste of the search process, Appendix A
examines two of the most promising candidate conjuncts, and shows
why they failed
to become part of the final invariant.

\section{Proof of correctness}
\label{sec:proof}

This section presents the proof of the theorem given in 
Section~\ref{sec:overview}:

{\it Theorem:} In any execution state, if there are
no subsequent join or fail events, then eventually the network will
become ideal and remain ideal.

As has been mentioned, the formal specification of correct Chord
is written in Alloy.
The Alloy language combines first-order predicate logic, relational
algebra, and transitive closure.
The Alloy Analyzer verifies properties by means of 
exhaustive enumeration of instances over a bounded domain.
This push-button analysis either yields a counterexample, or proves that
the property holds in the bounded domain.
The proof here is a hybrid, including both lemmas proved
automatically by the Alloy Analyzer and lemmas proved manually.
The reasons for using Alloy in this work, as well as its limitations,
are discussed in \cite{compare}.

\subsection{Modeling concurrency}
\label{sec:concurrency}

The formal model uses shared memory communication between nodes to
simulate queries.
An event is an atomic operation, executed by a single node and
altering only its own state, that may use the result of a single
query.
Concurrency has interleaving semantics.
Thus the interleaved events model local 
computations performed by
nodes between or after queries.

In the model, fail and rectify operations are independent events.
Joins correspond to two events at the same node:
\begin{enumerate}
\item
The node queries a known member for its current successor and
executes an event of type
\small
{\tt JoinLookup}
\normalsize
if it gets one.
\item
The node queries its current successor for a successor list
and executes an event of type
\small
{\tt Join}
\normalsize
if it gets one.
\end{enumerate}
A stabilize operation corresponds to one or two events at the 
same node:
\begin{enumerate}
\item
The node queries its first successor for a predecessor and successor
list, and executes an event of type
\small
{\tt StabilizeFromOldSuccessor}
\normalsize
if it gets them.
\item
Otherwise the node queries subsequent successors in its list as
above, until it succeeds in querying a live successor and
executing an event of type
\small
{\tt StabilizeFromOldSuccessor}.
\normalsize
\item
If the acquired predecessor appears to be a better first successor,
the node queries it for its successor list and executes an event of type
\small
{\tt StabilizeFromNewSuccessor}
\normalsize
if it gets the list.
\end{enumerate}

As the event types are modeled in the form of logical constraints, it is
necessary to use Alloy analysis to check that the constraints are
consistent, {\it i.e.,} that events of the types can exist or occur.
This has been done, as is shown in full at
\url{http://www2.research.att.com/~pamela/chord.html}.
All other proof steps are also included.

A 
\small
{\tt JoinLookup}
\normalsize
event establishes a precondition for its
subsequent
\small
{\tt Join}
\normalsize
event.
How can we be sure that the precondition still holds when the 
\small
{\tt Join}
\normalsize
event occurs, knowing that other events can occur
between this event pair?
The precondition is
\small
\begin{verbatim}
no b: Network.base | Between[ n, b, j.newSucc ]
\end{verbatim}
\normalsize
where
\small
{\tt no}
\normalsize
is a quantifier meaning $\neg \exists$,
\small
{\tt Network.base}
\normalsize
is the set of members of the stable base,
\small
{\tt j}
\normalsize
is the actual event of type
\small
{\tt Join}
\normalsize
(an Alloy object),
\small
{\tt n}
\normalsize
is the node executing
\small
{\tt j},
\normalsize
and
\small
{\tt j.newSucc}
\normalsize
is the new successor of
\small
{\tt n}.
\normalsize
The precondition says that
no member of the stable base lies between
\small
{\tt n}
\normalsize
and its new successor in identifier order.
No term of this condition is mutable or time-dependent, so interleaved
events cannot falsify it.
Here is a place where the assumption of a stable base plays a direct
role in the proof.

\subsection{Establishing the invariant}

The next step of the proof is to establish the inductive invariant,
named 
\small
{\tt Valid}
\normalsize
in the model,
by proving that it is preserved by events of every type.
For each event type such as 
\small
{\tt Fail},
\normalsize
there is a lemma such as
\small
{\tt FailPreservesValidity},
\normalsize
which says that if the network state is valid immediately before a fail
event, then it is valid immediately after the fail event.

The lemmas are proved automatically by the Alloy Analyzer.
Specifically, they are checked by exhaustive enumeration over all
possible networks with
$r \leq 3$ and $n \leq 9$, where $n$ is the number of nodes (including
ring members, appendage members, and non-members).

There are three reasons for believing that this bounded verification
is sufficient to count as a proof:
\begin{itemize}
\item
From Section~\ref{sec:disorder}, a successor list that is disordered
is also a successor list that, interpreted as a path around the ring,
``wraps around'' the ring.
As the ring grows larger, it becomes increasingly difficult for a
successor list built from interactions with neighbors to wrap around,
causing anomalies.
In this argument ``small'' and ``large'' are relative to $r$.
The exhaustive enumeration covers cases in which ring size is
$3r$.
\item
Ring structures have many symmetries.
For example, it has been proved by Emerson and Namjoshi that for all
properties of adjacent pairs of nodes, rings of size 4 are sufficient
to exhibit all counterexamples \cite{cutoff}.
This is not directly relevant because Chord's properties are global
rather than pairwise, but it does indicate that anomalies in rings 
occur when the rings are small.
\item
During the experience of model exploration with Alloy,
with $r = 2$,
many new behaviors were found
by increasing the number of nodes from 5 to 6, and no new behaviors were
ever found by increasing the number of nodes from 6 to 7.
Also, no new behaviors were found by increasing $r$ from 2 to 3.
This makes $r = 3$ and $n = 9$ seem like a safe limit.
\end{itemize}

It is also worth noting that Chord operations, when applied to a network
state that does not satisfy a sufficiently strong invariant, produce 
an astonishing variety of weird counterexamples, which the Alloy
Analyzer finds easily.
Given the predictable human
propensity to see what we want to see, Alloy analysis
is far more credible than a manual proof of invariance would be.
The only proof that would be more credible would be a formal proof
for networks of any size, checked by an automated theorem prover.

\subsection{Guaranteeing progress}

For each type of event that repairs the ring structure,
there is a predicate
\small
{\tt EffectiveEventTypeEnabled[n,t]}.
\normalsize
For a node 
\small
{\tt n}
\normalsize
and state timestamped
\small
{\tt t, EffectiveEventTypeEnabled[n,t]} 
\normalsize
is true if and only if at time
\small
{\tt t},
\normalsize
an event of that type can occur at 
\small
{\tt n},
\normalsize
and if it does occur
it will change the state of
\small
{\tt n}.
\normalsize

The definitions of these predicates must be checked for correctness.
For each predicate, this is done by proving a lemma that if the
predicate is true, the state is valid, and the event occurs, then
after the event the state of {\tt n} is different.

The purpose of these definitions is to use the Alloy Analyzer to
prove two crucial lemmas.
The predicate 
\small
{\tt NetworkIsImprovable}
\normalsize
is true whenever some effective
repair event is enabled:
\small
\begin{verbatim}
pred NetworkIsImprovable [t: Time] {
   (some n: Node | EffectiveSFOSenabled [n, t])
|| (some n: Node | EffectiveSFNSenabled [n, t])
|| (some n, newPrdc: Node | 
      EffectiveRectifyEnabled [n, newPrdc, t] )
}
\end{verbatim}
\normalsize
The predicate is used to assert that when the network is valid and not
ideal, it can be improved by an enabled repair event:
\small
\begin{verbatim}
assert ValidNetworkIsImprovable {
   all t: Time | 
      Valid[t] && ! Ideal[t] 
   => NetworkIsImprovable[t]  
}
\end{verbatim}
\normalsize
We assume that if a repair event is enabled, it will eventually be
scheduled and executed.  Furthermore, once a network has become ideal,
no executed repair event will change the state:
\small
\begin{verbatim}
assert IdealNetworkIsNotImprovable {
   all t: Time | 
      Valid[t] && Ideal[t] 
   => ! NetworkIsImprovable[t]  
}
\end{verbatim}
\normalsize
Together these lemmas establish that whenever a network is in a
non-ideal state, an effective repair event will eventually be executed
and change the state.
As with the invariant-preserving lemmas, they are proved by
exhaustive enumeration over all possible networks with
$r \leq 3$ and $n \leq 9$.

The final step is to show that a sequence of effective repair events
must eventually terminate by making the network ideal. 
This step is informal.
We define a measure of the error in a Chord network, such that the 
measure of an ideal network is 0 and the measure of a non-ideal network
is a positive integer.
We will also show that
every effective repair event reduces the measure.
This will complete the proof
that a network with no new joins or fails will eventually
become ideal.

Let $s$ be the current size of the network (number of members).
This number is only changed by join and fail operations, and not by
any repair operations, so it remains the same throughout a repair-only
phase as hypothesized by the theorem.
The error of a pointer is defined as follows:
\begin{itemize}
\item
The error of a predecessor or first successor is 0 if it points to the
globally correct member (in the sense of identifier order), 1 if it
points to the next-most-correct member, . . . $s - 1$ if it points to
the least globally correct member, 
$s$ if there is no pointer
(possible only for a predecessor), and
$s + 1$ if it points to a non-member.
\item
The error of a second or later successor is 0 if its node's successor
is live and the pointer
matches the corresponding pointer of its node's successor's
successor list.
This holds regardless of whether the value of the pointer is globally
correct or not.
The error of the second or later successor is 1 otherwise.
\end{itemize}
The total error or just ``error'' of a network
is defined as the sum over all members and
all pointers of the pointer error.

We now explain the effect of each repair event on the network error.
First, there are two cases of effective
\small
{\tt StabilizeFromOldSuccessor}
\normalsize
events (see Section~\ref{sec:concurrency}).
\begin{enumerate}
\item
In one case, the member's old successor was dead and is replaced by
a live successor.
In this case the error of the member's first successor changes from
$s + 1$ to something less than $s$.
The error of its second and later successors changes from 1 to 0.
\item
In the other case, the member's old successor was live.
In this case the error of the member's first successor does not change,
but the error of at least some
of its second and later successors changes from 1 to 0.
Note that if the stabilizing member is completely up-to-date and
no part of its successor list changes, then this is not an effective
event, and need not be considered.
\end{enumerate}

An effective 
\small
{\tt StabilizeFromNewSuccessor}
\normalsize
always
reduces the error of the first successor.  
After the event the error of all second and later successors is 0,
so it may be decreased and is not increased.

There are three cases of effective
\small
{\tt Rectify}
\normalsize
events (see Section~\ref{sec:spec}).
\begin{enumerate}
\item
In one case there was no previous predecessor, and the error of
the predecessor changes from $s$ to something less than $s$.
\item
In another case,
the previous predecessor was dead.
In this case the error of
the predecessor changes from $s + 1$ to something less than $s$.
\item
In the third case,
the previous predecessor was live.
In this case the error of
the predecessor is always reduced.
\end{enumerate}
In each case, for each event type, the error is reduced by the event.
$\Box$

\section{Conclusion}

The basic design of the Chord ring-maintenance protocol is 
extraordinary in its achievement of consistency with so little
overhead, so little synchronization, and such weak assumptions of
fairness.

Although refining the design and proving it correct
was difficult, modeling the original version of Chord and using
the Alloy Analyzer to check whether it satisfied its claims was not
difficult.
Alloy and other similar tools such as model-checkers have become
mature in the years since Chord was originally designed, and some
such tool should be a part of every protocol designer's toolkit.

As practical consequences of this work, new implementers of
Chord should use the specifications in this
paper.
Nodes that fail and restart can rejoin a Chord network immediately
on restart.

Previous implementers of Chord should
check their implementations of operations
for bugs.
They should introduce procedures for better initialization,
or at least global monitoring that the ring grows to a safe size without
anomalies.
After this initial phase, only local checking of invariants on
extended successor lists is advisable for reliability and security.
Designers and implementers of other ring-shaped distributed
data structures should consider what their invariants are,
how they are related to Chord's, and how they might be monitored.

On the theoretical side,
the most interesting future work would be to attempt to prove,
with the same standard of formality, the
correctness of enhancements such as
protection against malicious peers
\cite{awerbuch-robust,chord-byz,sechord},
key consistency and data consistency \cite{scatter},
range queries \cite{rangequeries},
and atomic access to replicated data \cite{atomicchord,etna}.
Although some of these enhancements use probabilistic reasoning
more heavily, 
probabilistic model-checking and verification are now coming into
their own.

\section*{Acknowledgments}

Helpful discussions with
Bharath Balasubramanian,
Ernie Cohen,
Patrick Cousot,
Gerard Holzmann,
Daniel Jackson,
Arvind Krishnamurthy,
Gary Leavens,
Pete Manolios,
Annabelle McIver,
Jay Misra,
Andreas Podelski,
Emina Torlak,
Natarajan Shankar, and
Jim Woodcock
have contributed greatly to this work.

\bibliographystyle{IEEEtran}
\bibliography{correct}

\section*{Appendix A: Investigation of trial invariants}

\begin{figure*}
\centering
\includegraphics[scale=0.80]{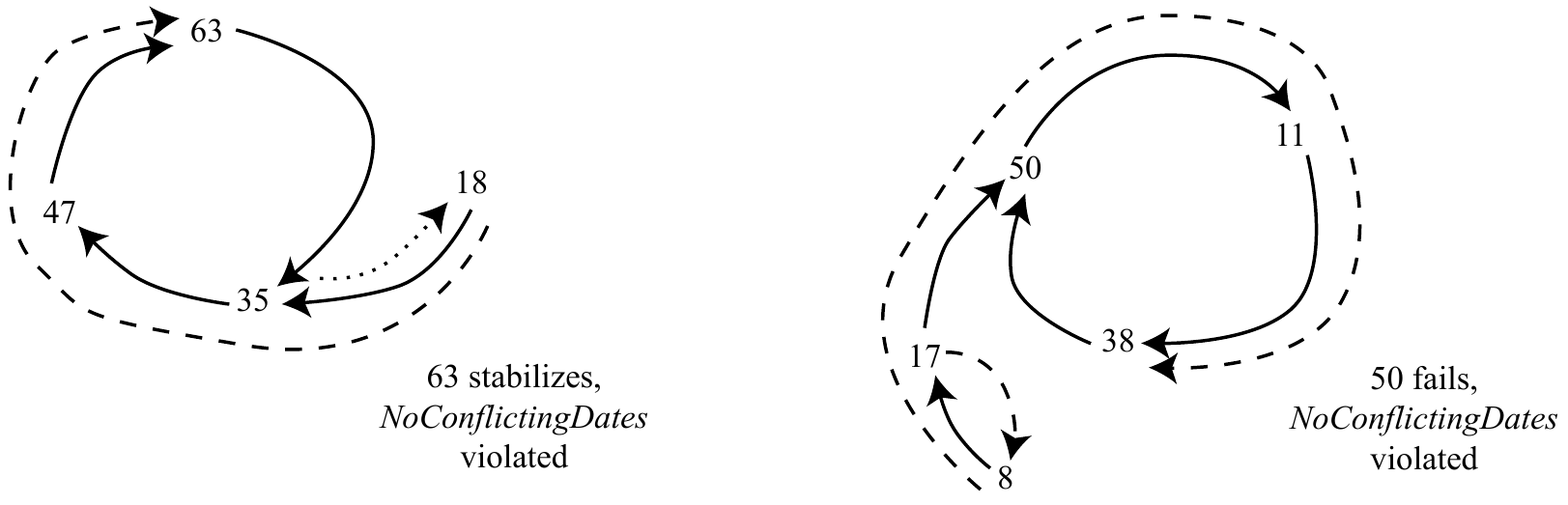}
\caption{Two counterexamples to a trial inductive invariant.
Second successors not drawn are correct, {\it i.e.,} they are the
successors of the nodes' successors.}
\label{fig:ncd}
\end{figure*}

Section~\ref{sec:disorder} explained that a member {\it skips} another
member if the skipping member's extended successor list does not
mention the skipped member, yet the skipped member fits between a pair
that is in the list.
The intuition is that the skipped member is too new to be known to
the skipping member.
Based on this intuition, we can define a predicate {\it MustPreDate(n1,n2)}
that is true if and only if {\it n1} and {\it n2} are ring members,
and there is a third ring member {\it n3} (possibly the same as {\it n1})
that mentions {\it n1} and skips {\it n2}.

One of the most promising candidate conjuncts for the inductive
invariant was {\it NoConflictingDates}, which says that there is no
pair {\it (n1,n2)} such that {\it MustPreDate(n1,n2)} and
{\it MustPreDate(n2,n1)}.

{\it NoConflictingDates} is closely related to another candidate
conjunct, {\it NoEjects}.
{\it NoEjects} simply says that a member in the ring has no
successor in its list that points to an appendage.
For example, in Figure~\ref{fig:wrap}, ring member 52 has eject 45.

The two are related in the sense that violations of each can cause
violations of the other.
First, consider an arc of a ring {\it (v, w, x, y, z)}
where {\it v} skips {\it x} because its second successor is {\it y},
and {\it w} skips {\it y} because its second successor is {\it z}.
Note that this violates {\it NoConflictingDates}:
Because 
{\it v} skips {\it x}, all of {\it v, w,} and {\it y} must pre-date
{\it x}.
Because 
{\it w} skips {\it y}, all of {\it w, x,} and {\it z} must pre-date
{\it y}.
Now if {\it x} fails, then the arc of best successors becomes
{\it (v, w, z)}, with {\it y} as an appendage connected to the ring
at {\it z}.
Now the second successor of {\it v} is an eject, pointing outside
the ring to {\it y}.

Figure~\ref{fig:wrap} is an example that goes in the other direction.
It begins with a violation of
{\it NoEjects}, and ends with a violation of
{\it NoConflictingDates} (45 pre-dates 52 and 52 pre-dates 45).

This raises the enticing possibility that the four original
conjuncts, plus {\it NoDuplicates, OrderedSuccessorLists, 
NoConflictingDates,} and {\it NoEjects}, might make an inductive
invariant.
Unfortunately it is not, and
Figure~\ref{fig:ncd} shows two separate counterexamples
with $r = 2$.
Both sides of the figure satisfy the proposed invariant.
Yet if 63 stabilizes on the left side, 
{\it NoConflictingDates} is violated (18 pre-dates 47 and 47 pre-dates
18).
Also, if 50 fails on the right side,
{\it NoConflictingDates} is violated (each pair in 11, 17, and 38 has
a date conflict).

At this stage of the investigation there are two possibilities:
\begin{itemize}
\item
The proposed invariant is simply not an invariant.
It is worth noting that the inductive invariant with {\it BaseNotSkipped}
does not imply either {\it NoConflictingDates} or
{\it NoEjects}, although it does imply {\it NoDuplicates} and
{\it OrderedSuccessorLists}.
\item
The proposed invariant is a true invariant of Chord, and
the networks in Figure~\ref{fig:ncd} are false counterexamples
because they could never occur during Chord operation.
They do look weird!
If so, however, the proposed invariant is not inductive.
To make it inductive we must add {\it other as-yet-unknown
conjuncts} to exclude the networks in the figure.
\end{itemize}
There is no way to know which possibility is the true one, and
either way we are no closer to a final inductive invariant.

It may seem that it would be easy to add conjuncts to exclude
the networks in Figure~\ref{fig:ncd}, but looks are deceiving.
Every conjunct that makes the problem easier by excluding 
more pre-operation
states also makes the problem harder by excluding more
post-operation states and thus generating new counterexamples.
The process does not converge.
For those who would like to experiment on their own,
\url{http://www2.research.att.com/~pamela/chordnobase.als}
is a model of Chord without a stable base that can be used
conveniently for this purpose.

To explore in a different direction,
it would be satisfying to achieve certainty on the feasibility
of the networks in Figure~\ref{fig:ncd} by 
using a model-checker to generate the entire reachable state space of a
Chord network with $r = 2$, for some $n > 5$.
(A network might be reachable with the past participation
of other nodes that are no longer members, hence the need for more
nodes.)
This has been attempted with the model-checker Spin \cite{spin}, but
the analysis is too computationally complex.
Even analysis of a simpler Chord model with restricted concurrency
did not reach the entire state space with $n = 5$ \cite{compare}.

\end{document}